\documentclass[aps,pra,preprint,groupedaddress,showpacs]{revtex4}
\usepackage{epsfig}

\begin{document}

\title{Efficient positronium laser excitation for antihydrogen production
in a magnetic field}
\author{F. Castelli, I. Boscolo, S. Cialdi, M.G. Giammarchi,}
\affiliation{INFN and Universit\`a di Milano, via Celoria 16, 20133 Milano, Italy}
\author{D. Comparat}
\affiliation{Laboratoire Aim\'e Cotton, CNRS Univ Paris-Sud B\^{a}t.505, 91405 Orsay, France}

\date{\today}

\begin{abstract} Antihydrogen production by charge exchange reaction between
  Positronium (Ps) atoms and antiprotons requires an efficient excitation of
  Ps atoms up to high-$n$ levels (Rydberg levels).  In this study it is
  assumed that a Ps cloud is produced within a relatively strong uniform
  magnetic field (1 Tesla) and with a relatively high temperature (100 K).
  Consequently, the structure of energy levels are deeply modified
  by Zeeman and motional Stark effects. A two--step laser light excitation,
  the first one from ground to $n=3$ and the second from this level to a Rydberg
  level, is proposed and the physics of the problem is discussed.  We derive
  a simple formula giving the absorption probability with substantially
  incoherent laser pulses.  A  30\% population deposition in high-$n$ states
  can be reached with feasible lasers suitably tailored in power and spectral
  bandwidth.
  \end{abstract}


\maketitle

\section{Introduction}
Some fundamental questions of modern physics relevant to the unification of
gravity with the other fundamental interactions such as models involving vector and
scalar gravitons and matter anti-matter symmetry (CPT) can be enlightened via
experiments with antimatter \cite{hughes}. In particular, some quantum
gravity models claim for possible violations of the equivalence principle of
General Relativity in antimatter \cite{nieto}. Testing the validity of this
principle is an important issue, and may involve measurements
of the inertial and gravitational mass equality using different experimental
settings, for example determining gravitational acceleration on cold atoms
\cite{peters}.  The authors of the present paper are co-proponents of the
AEGIS (Antimatter Experiment: Gravity, Interferometry, Spectroscopy)
experiment \cite{kellerbauer,aegis} aiming to measure the Earth Gravitational
acceleration $\bar{g}$ on a cold and collimated antihydrogen $\bar{H}$ beam.

The production of cold antihydrogen bunches occurs in the charge transfer of
a cloud of Rydberg excited positronium atoms (Ps) (stored in a
magnetic trap) with a bunch of cold antiproton $\bar{p}$ by means of the
reaction $Ps^* + \overline{p} \rightarrow \overline{H} + e^-$ \cite{hessels}.
Ps atoms and antiprotons are prepared inside a cryostat.  Here we will focus
on the excitation at high quantum levels of the Ps atomic cloud.

The $\bar{H}$ formation process has to be very efficient. The number
antihydrogen atoms produced in the charge exchange reaction is expressed,
with obvious notation, as $N_{\bar{H}} = \rho \, N_{Ps}\, N_{\bar{p}} \,
\sigma / A$ where $\rho$ is the overlap factor between the trapped
${\bar{p}}$ and the Ps cloud with transverse area $A$. Since the cross
section $\sigma$ depends on the fourth power of the principal quantum
number $n$ of the excited Ps \cite{hessels} ($\sigma \propto n^4 \pi a_0^2$,
where $a_0$ is the Bohr radius), $n$ can be chosen to be in the range from $20$ to
$30$. As we shall see later, higher $n$ values should be avoided in order to reduce  ionization losses due to stray fields and dipole-dipole interactions. Incidentally, the higher the $n$-value the longer the Ps lifetime.

Positronium excitation to these high-$n$ levels (the so-called Rydberg
levels) can be obtained either via collisions or via photon excitation. In
reference \cite{hessels} Ps excitation was proposed and tested through Cs
excitation by light and a successive charge exchange reaction with
positrons. We propose (in the AEGIS experiment) a direct Ps excitation by a two
step light excitation using two simultaneous laser pulses with different
wavelengths. This excitation process should be very efficient, more controllable and of
simpler and more compact experimental set up than the previous one.

An efficient way of producing Ps atoms is by letting positrons impinge on a
proper porous silica surface \cite{gidley}. The Ps exiting the target surface
forms an expanding cloud at a temperature up to 100 K, and the cloud is
constrained by a magnetic trap with a relatively strong field $\vec{B}$ of
about 1 Tesla \cite{kellerbauer}. Ps atom resonances will then be broadened by
Doppler effect because the atoms have random velocities of the order $v \sim
10^5\, m/s$ at the 100 K reference temperature. Moreover the sublevels of
a Rydberg excited state will be mixed and separated in energy by the motional
Stark effect and by linear and quadratic Zeeman splitting. Because of these
effects the transition will be from ground or from
a low--excited level to a broadened Rydberg
level--band. This last is in fact the relevant characteristic which
distinguishes the Ps laser excitation from the usual atomic excitation to
Rydberg levels; therefore it requires a careful theoretical analysis and a suitable
experimental setup.  The characteristics of laser pulses in terms of power
and spectral bandwidth must be tailored to the geometry, the Rydberg
level--band and the timing of the Ps expanding cloud. The laser power
must be enough to ensure the whole Ps cloud excitation within a few
nanoseconds.

The interest on laser Ps excitation started with theoretical studies on the delayed annihilation \cite{faisal} induced by excitation towards $p$ and $d$ states using nanosecond or longer pulses. Subpicosecond pulses of very intense laser radiation were proposed and analyzed both for spectroscopic studies and for antihydrogen formation by the charge exchange process \cite{madsen}, calculating population deposition on low energy states essentially by two-photon absorption. Generally, these studies did not consider Rydberg states, nor the presence of a magnetic field. The problem of Ps Rydberg excitation with nanosecond laser pulses in a magnetic environment was faced and experimentally performed in Ref. \cite{ziock} for $n$ up to 19, but in a different regime with respect to ours,
as discussed below.

The photo--excitation of Ps to the Rydberg band requires photon energies
close to 6.8 eV. Laser systems at the corresponding wavelength ($\approx 180$
nm) are not commercially available, hence a two--step excitation was
required. We are taking into consideration the transition from the ground
state to $n=3$ state ($\lambda=205$ nm), and then to high--$n$ levels
($\lambda \sim 1670$ nm). This choice seems more adequate than the other
possible two step sequence $1 \rightarrow 2$ and $2 \rightarrow$ high-$n$
because the level $n=2$ has a three time shorter lifetime than
the the $n=3$ level (3 ns against 10.5 ns) and, in addition, population loss
becomes dynamically relevant. The two photon excitation $1 \rightarrow 2
\rightarrow n $ was reported in Ref.~\cite{ziock}.

A commercial Dye--prism laser (optically pumped by the second harmonic of a
Q--switched Nd:YAG laser) coupled to a third harmonic generator can provide
the 205 nm photons for the first transition. The laser for the second
transition is yet to be developed and it is being proposed for the AEGIS experiment
\cite{kellerbauer}.  In the discussed two--photon excitation we must use a
relatively high intensity pulse in order to have an efficient transition
process. Since one needs to avoid losses in the excited population due to the
short lifetime of the intermediate levels, the pulse time length cannot
exceed a few nanoseconds; their expected value is around 5 ns.

We perform the calculation of the excitation process to relate the absorption probability rate to the laser power and its spectral bandwidth characteristics, using a suitable definition of transition saturation fluence.

\section{Modeling Ps excitation from $n = 1$ to high--$n$ levels}

Here we consider a simple theoretical model of Ps excitation to
calculate laser saturation fluence and useful bandwidth. The
excitation of Ps in high--$n$ state is described as a cascade or a
two-step transition: a first step by a resonant excitation from $n=1$
to $n=3$, and a second step by a near resonant excitation from $n=3$
to high--$n$. The spectral profile of the two laser intensities is
characterized by a Gaussian function whose full width at half maximum (FWHM) $\Delta \lambda_L$ is matched to a selected Rydberg level--bandwidth around a
definite $n$ state. The broad laser linewidths come along with a
coherence time $\Delta t_ {coh} = \lambda^2 / c \, \Delta\lambda_L$,
where $\lambda$ is the central wavelength of the proper
transition. This parameter turns out to be up to three orders of magnitude shorter
than the average 5 ns duration of the laser pulses \cite{kellerbauer},
hence we are operating
with a completely incoherent excitation for both transitions.

The detailed structure of the optical transition to high--$n$ energy levels of
positronium is dominated by:
\begin{itemize}
    \item the Doppler effect;
    \item the Zeeman and motional Stark effects,
\end{itemize}
because their energy contribution is larger than hyperfine and spin--orbit
splitting in the experimental conditions. However, the importance of these
effects for the effective level structure is completely different for $n = 3$ and for
Rydberg states. As we shall see, while the first transition is marginally
affected by Zeeman and Stark effects, the high $n$ levels involved in the
second transition are turned into energy bands by Stark effect, strongly
affecting the physics of the excitation. Therefore we will treat
the two transitions separately.

The challenging problem of Rydberg states of an atom moving in a magnetic field
has attracted many experimental and
theoretical researches \cite{main}. The Ps atom is a special case
because the first order Zeeman effect, \emph{i.e.} the direct interaction
between magnetic field and magnetic dipoles, only mixes ortho and para Ps
states with $m_S = 0$ without affecting orbital quantum numbers
\cite{feinberg}.  This interaction energy contribution amounts to $\pm 1.2
\times 10^{-4}$ eV for $B = 1$~T (much lower than the actual Doppler
broadening, see below), while the energy of $m_S = \pm 1$ states remain
unchanged. The level mixing leads to the well known enhancement of the
average annihilation rate of the Ps thermal ground state $n = 1$ \cite{rich},
leaving in fact only the ortho--Ps states with $m_S = \pm 1$ surviving in the
Ps cloud expanding from the silica converter. From the observation that the
electric dipole selection rules for optical transitions impose conservation
of spin quantum number, and that the broadband characteristics of our laser
overlaps the Zeeman splitting, we may conclude that in first approximation
this effect does not play any role in the transition. Thus we will
concentrate our attention only on the dominant motional Stark effect. As a
final note we observe that the quadratic (diamagnetic) Zeeman effect can be
discarded because it gives an energy contribution only for higher magnetic
fields and high $n$ (being proportional to $n^4$) \cite{feinberg,garstang},
hence in a regime where the motional Stark effect dictates the transition
structure.

Since Doppler and Stark effects depend on temperature, in the following
calculations we select for definiteness the reference temperature of 100 K,
which corresponds to the largest Ps cloud exiting the converter, and
consequently greater laser powers. This choice ensures a successful Ps
excitation with lower temperatures as well.

\subsection{Excitation from $n = 1$ to $n = 3$}

In the first excitation step the Doppler linewidth $\Delta\lambda_D$,
scaling as $\sqrt{T}$, turns out to be around $4.4 \times 10^{-2}$ nm,
corresponding to the energy broadening $\Delta E_D \simeq 1.3 \times 10^{-3}$
eV. A motional Stark electric field $\vec{E} = \vec{v} \times \vec{B}$, where $\vec{B}$ is taken for definiteness along the $z$ axis, is
induced by the Ps motion within the relatively strong magnetic field $B \sim
1$ T \cite{dermer} . This effect splits the sub--levels of the
state $n = 3$ in energy and leads to some mixing of quantum numbers $m$ and $\ell$, due
to the breaking of the axial symmetry of moving Ps atoms. The maximum
broadening due to this effect is evaluated as $\Delta \lambda_S \simeq 1.8
\times 10^{-3}$ nm (the total energy width of the sub--level structure amount
to $\Delta E_S \simeq 5.3 \times 10^{-5}$ eV) \cite{note1}, negligible with
respect to the Doppler broadening. Therefore we conclude that the width of
the transition $1 \rightarrow 3$ is dominated by the Doppler broadening, and
the laser linewidth must be provided accordingly.

Since Ps excitation is incoherent, the saturation fluence is
calculated by a rate equation model (see the Appendix). The excitation
probability for unit time is
\begin{equation}\label{excp1}
    W_{13}(t) \, = \, \int d\omega \, \frac{I(\omega,t)}{\hbar \, \omega}
    \, \, \sigma_{13}(\omega)
\end{equation}
where $I(\omega,t)$ is the power spectrum of the laser pulse and
the absorption cross section $\sigma_{13}$ is
\begin{equation}\label{cs1}
    \sigma_{13}(\omega) \, = \, \frac{\hbar \, \omega}{c} \,
    g_D(\omega - \omega_{13}) \, B_{1 \rightarrow 3}(\omega)
\end{equation}
($c$ is the vacuum light speed and $\hbar$ is the Planck constant divided by $2 \pi$). The function $g_D(\omega - \omega_{13})$ is the normalized lineshape representing
the Doppler broadened line, \emph{i.e.} a Gaussian function centered on the
transition frequency $\omega_{13}$ and with a FWHM
corresponding to the Doppler linewidth $\Delta\lambda_D $. The factor $B_{1
  \rightarrow 3}(\omega)$ is the absorption Einstein coefficient appropriate
to the dipole--allowed transition (the frequency dependence is inserted for
consistency with the following subsection). In first approximation
\cite{dermer} this coefficient coincides with that of the unperturbed
transition $(1,0,0) \rightarrow (3,1,m)$ where $m$ can be selected by the
laser polarization, specified by the unit vector $\vec{\epsilon}$. Hence one
has $B_{1 \rightarrow 3}(\omega) = |d_{1 \rightarrow 3}|^2 \pi /
\varepsilon_0 \hbar^2$ with the electric dipole matrix element $d_{1
  \rightarrow 3} = \langle\psi_{100} | e \, \vec{r} \cdot \vec{\epsilon} \,
|\psi_{31m} \rangle$ (where $e$ is the electron charge and $\vec{r}$ the
position operator) calculated using standard methods from the general theory
of radiative transition in atomic physics \cite{sobelman} and the exact
Gordon formula \cite{gordon}.

By matching the resonant laser  linewidth to the Doppler broadening
(aiming to maximize Ps cloud covering in the spectral domain)
and assuming for definiteness linear laser polarization along the $z$ axis,
from the results of the rate equation theory developed in Appendix
(see Eq.~(\ref{fsat})) we can determine the saturation fluence for the
first transition as
\begin{equation}\label{f1}
    F_{sat}(1 \rightarrow 3) =  \frac{c^2}{B_{1 \rightarrow 3}}
    \sqrt{\frac{2 \, \pi^3}{\ln 2}} \, \cdot \,
    \frac{\Delta\lambda_D}{\lambda_{13}^2}  =  93.3 \,
    \mu\mbox{J/cm}^2
\end{equation}
This equation gives the lowest pulse fluence needed for reaching transition
saturation. The energy of the exciting laser pulse will depend on the laser
spot-size (which must overlap the Ps cloud). Assuming a transverse cloud FWHM
dimension of $\Delta r =2.8\,$ mm (the Ps cloud section of 6 mm$^2$ of AEGIS
proposal) and fixing the fluence value $F_0$ at the maximum of the transverse
Gaussian laser profile as $F_0 = 2 F_{sat}$, the laser pulse energy ($E = \pi
(F_0/2) (\Delta r / 1.177)^2$) comes out to be $E_I = 16.2 \, \mu$J.

\subsection{Excitation from $n = 3$ to Rydberg levels}

\begin{figure}[!ht]
\center
\includegraphics[width=8cm,clip=true]{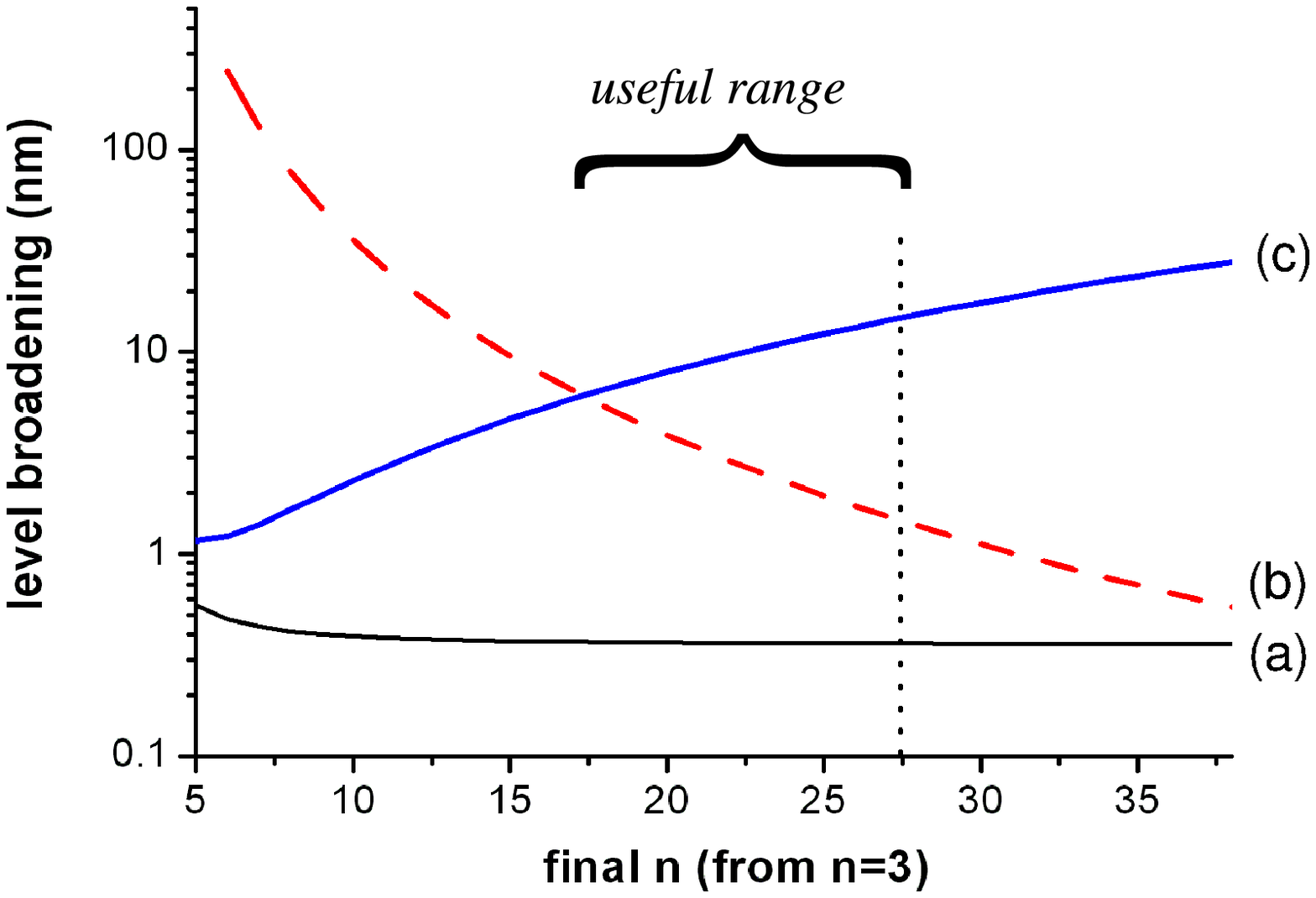}
\caption{(Color online) Doppler (a) and Stark (c) line--broadenings as a function of
the principal quantum number $n$ for the transition $3 \rightarrow
n$. The dashed line (b) shows the energy distance (in nm) between adjacent
unperturbed $n$ states. The dotted vertical line is the ionization
limit for the lowest sub level. The useful range for Ps Rydberg
excitation is indicated.}
\label{stark-graph}
\end{figure}

The physics of the second transition $n=3 \rightarrow$ high--$n$ is
significantly different. The Doppler broadening is practically
independent of $n$ and turns out to be around 0.35 nm at 100 K
(corresponding to an energy broadening of $1.6 \times 10^{-4}$ eV),
whereas the Stark motional effect turns out to be many times higher,
as depicted in Fig.~\ref{stark-graph}.

The effect due to the motional Stark electric field becomes the dominant
characteristic of the transition. Because of it, the degenerate high--$n$
levels become fans or manifolds of their $n^2$ sub--levels
with a complete mixing of their $m$ and $\ell$ sub states, while the mixing
between $n$--levels in Positronium atoms does not occur to a good extent
\cite{gallag}. Owing to the $m$ and $\ell$ sub--level mixing, these
unperturbed quantum numbers are no longer good quantum numbers labelling the
states, at variance with the principal quantum number $n$ which retains its
role \cite{dermer,ziock}. The energy width $\Delta E_S$ of a fan can be
evaluated from the usual theory of the Stark effect, and increases both with
the magnetic field and $n$ as
\begin{eqnarray}
  \Delta E_S \, &=& \, 6 \, e \, a_0 \, n \left( n-1 \right) \,
    |\vec{E}(v_\perp)| \nonumber \\
  &=& \, 6 \, e \, a_0 \, n \left( n-1 \right)
    \, B \, \sqrt{k_B T / m}    \label{estark}
\end{eqnarray}
where $a_0$ is the Bohr radius, $v_\perp =\sqrt{k_B T / m}\,$ is the
positronium atom thermal transverse velocity ($m$ being the positronium mass)
and a factor of $2$ comes from the fact that the radius of the ground state
of a Ps atom is equal to $2 a_0$. The broadening $\Delta\lambda_S \simeq
\Delta E_S \, \lambda^2 / 2 \pi c \hbar$ of the transition is shown in
Fig.~\ref{stark-graph}. It is worth noting that the splitting between
adjacent unperturbed energy levels (which energy is $E_n = - 13.6 \, \mbox{eV}/
2 n^2$) decreases with $n$
\begin{equation}\label{esplit}
    \Delta E_n \, \simeq \, 13.6 \, \mbox{eV} \, \cdot \frac{1}{n^3}
\end{equation}
as shown in the figure. Therefore for $n > 16$ the bandwidth filled by the
sublevels relative to an $n$ state becomes overwhelmingly greater than the
interval between two adjacent $n$--levels.  Thus, at $n$ larger than 16 an
interleaving of many sublevels is expected.  The range of $n$ levels useful
for the charge transfer reaction and efficient $\bar{H}$ formation starts
from $n \sim 20$, i.e. in the region of notable level mixing.
We would like to remark that our calculations refer to the cases
where $\Delta E_S \gg \Delta E_n$ (as it comes
out from Fig. \ref{stark-graph}).

Another effect of the motional Stark electric field that has to be considered
is the possible atom ionization: the transition from the bound state to an
ionized state occurs from the bottom sub--level of an $n$--fan (the
\emph{red--state} \cite{gallag}) to the unbound states. This action
determines an upper $n$--level useful for our purpose. The minimum Stark
electric field $\vec{E}_{min}$ which induces high ionization probability at
the lowest energy $E = E_n - 3 e a_0 n \left( n-1 \right) |\vec{E}_{min}|$ of
the level fan is calculated as \cite{gallag}
\begin{equation}\label{eion}
    |\vec{E}_{min}| \, = \, \frac{e}{16 \, \pi \, \varepsilon_0 \, a_0^2}
    \,\,\, \frac{1}{9 \, n^4} \,\, .
\end{equation}
Hence, for $B = 1$ T and for the reference temperature of 100~K, the
ionization starts affecting part of the level fan for $n > 27$. This
ionization limit, and the useful range for $n$, are indicated in
Fig.~\ref{stark-graph}.

\begin{figure}[!ht]
\center
\includegraphics[width=8.1cm,clip=true]{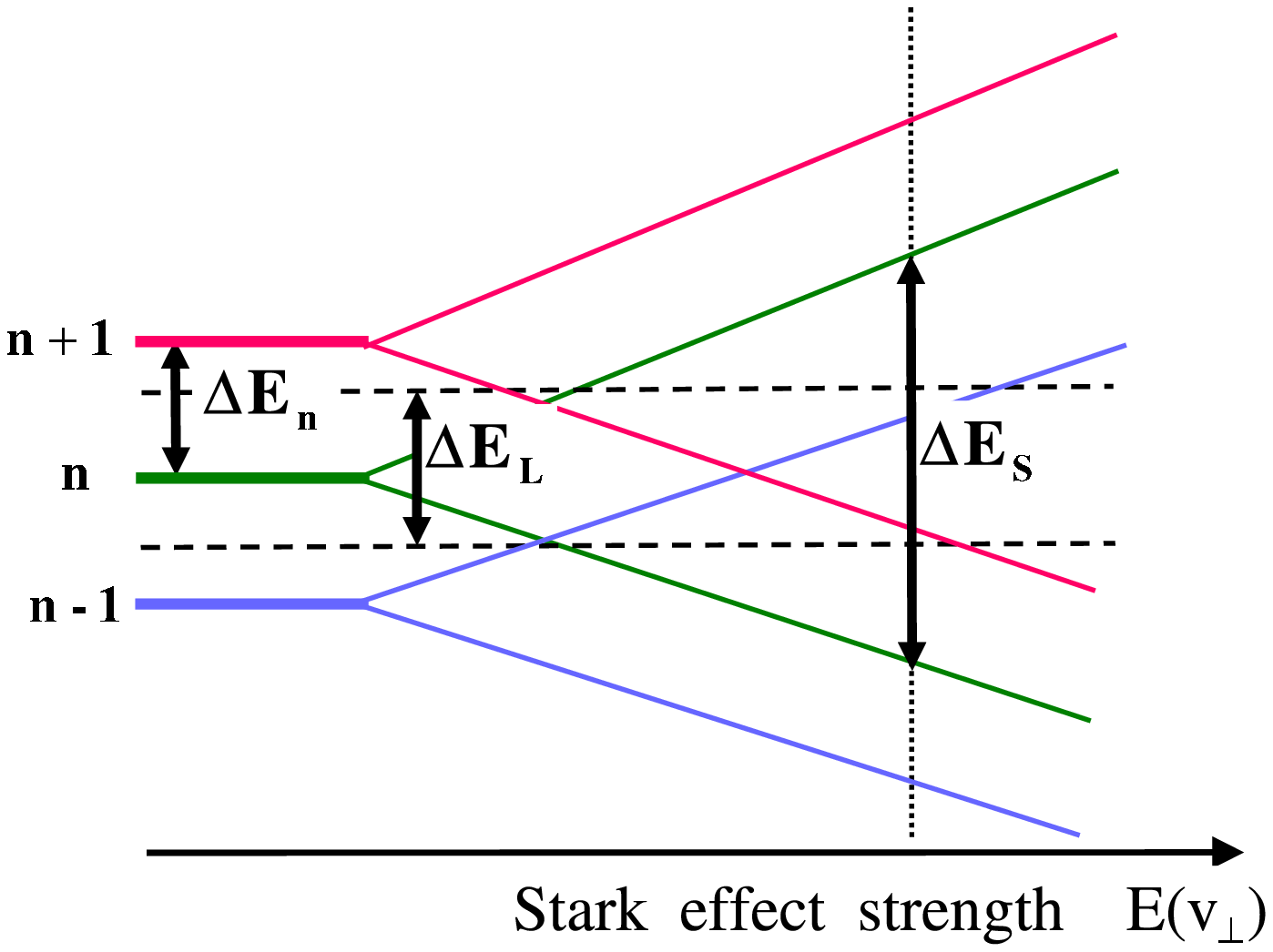}
\caption{(Color online) Schematic of the level mixing with respect to the laser energy bandwidth $\Delta E_L$, as a function of the strength of the motional Stark
electric field ($v_\perp$ is the Ps transverse velocity). The initial $n$-level is $n^2$ times degenerate, and its energy distance with the adjacent unperturbed level is $\Delta E_n$. The degeneracy is lifted by the Stark effect and the energy width of the sublevel fan is $\Delta E_S$.}
\label{figlivelli}
\end{figure}

Let us consider the distribution of the sublevels
to calculate the $3 \rightarrow$ high--$n$ transition
probability. For a given $n$, a uniform distribution of the $n^2$ fan sublevels within
the motional Stark energy width $\Delta E_S$ may be assumed.
The number of originally unperturbed $n$--levels interleaved with that
reference $n$--level within its fan width is approximately (see also
Fig.~\ref{figlivelli})
\begin{equation}\label{nn}
    N_n \, \simeq \, \frac{\Delta E_S}{\Delta E_n} \, \simeq \,
    n^5 \, \frac{6 \, e \, a_0}{13.6 \, \mbox{eV}} \, |\vec{E}(v_\perp)| \, .
\end{equation}
Therefore the sub level density per unit angular frequency results in
\begin{equation}\label{density}
    \rho(\omega) \, = \, \frac{n^2 \, N_n}{\Delta E_S/ \hbar} \, = \,
    n^2  \cdot \frac{\hbar}{\Delta E_n} \, = \,
    n^5 \, \frac{\hbar}{13.6 \; eV} \, ,
 \end{equation}
independent of the induced Stark field and consequently on the positron
velocity. We stress that this density occurs with a motional Stark effect
high enough (that is a transverse positronium velocity in a high magnetic
field) for producing an interleaving of many $n$-level fans, and it
increases very fast with $n$. Within the uninteresting region of the
transitions to $n <16$ it is easy to see that the density of sublevels is a
constant on $n$.

The bandwidth $\Delta\lambda_L$ of the second transition laser has to be
wider than the Doppler bandwidth $\Delta\lambda_D$ (for Ps cloud efficient
covering), and also narrower than $\Delta \lambda_S$ for constraining Ps atom
excitation within a reasonable narrow energy band (seeming this suitable for an
efficient charge transfer reaction). Therefore the laser energy bandwidth
$\Delta E_L$ is selected to be smaller than $\Delta E_S$ so that the sublevel density of the
above equation holds.  In these conditions, all the mixed sublevels with
transition energy under the laser bandwidth can be populated, at variance
with those foreseen by the electric dipole selection rule \cite{ziock}.

In Fig. \ref{figlivelli} a schematic picture of the level mixing is shown.
The number of levels per unit bandwidth remains, in a crude approximation,
unchanged with the increase of the fan aperture because the sublevels lost at
the border of the initially chosen laser bandwidth $\Delta E_L$ are compensated
by the arrival of sublevels coming form the nearby $n$-states.

The incoherent excitation probability per unit time of the transition
$3 \rightarrow$ high--$n$ is
\begin{equation}\label{excp2} W_{3n}(t) \, = \,
    \int_{\Delta E_L} d\omega \, \frac{I(\omega,t)}{\hbar \, \omega}
    \, \, \sigma_{3n}(\omega)
\end{equation}
The absorption cross section  $\sigma_{3n}(\omega)$, in this connection, can
be recast as
\begin{equation}\label{cs}
    \sigma_{3n}(\omega) \, = \, \frac{\hbar \, \omega}{c} \, \rho(\omega)
    \, B_S(\omega).
\end{equation}
with the absorption coefficient $B_S(\omega)$ appropriate for the excitation of a
single sublevel in the quasi--continuum Rydberg level band. By definition
this coefficient must be proportional to the square modulus of the electric dipole
matrix element:
\begin{equation}
    B_S(\omega) \, \propto \, |\langle\psi_{n\alpha} |e \, \vec{r} \cdot \vec{\epsilon}
    \, |\psi_{31m'} \rangle|^2
\end{equation}
where $\psi_{n\alpha}$ is the wavefunction of a Rydberg sublevel (with
$\ell$ and $m$ mixed) connected by the transition energy $\hbar\omega$
with the low level $\psi_{31m'}$ which is assumed excited by the first
laser. The following considerations allows us to estimate magnitude of
$B_S(\omega)$. The wavefunction $\psi_{n\alpha}$ relative to the
sublevel $n\alpha$ is given by a linear superposition of the $n^2$
unperturbed wavefunctions with suitable coefficients
\begin{equation}
 \psi_{n\alpha}= \sum_{lm} c_{lm} \, \psi_{nlm}.
\end{equation}
From the normalization condition and assuming a large spreading of
$\psi_{n\alpha}$ over the $\psi_{nlm}$, we get $|c_{lm}| \simeq
1/n$. Using the electric dipole selection rules, which select the
final state $nlm$, we obtain a simple formula connecting $B_S(\omega)$
with the absorption Einstein coefficient for the unperturbed $3 \rightarrow$
high--$n$ transition
\begin{eqnarray}
  B_S(\omega) \, & \propto & \, \frac{1}{n^2} \,
  |\langle\psi_{nlm} |e \,\vec{r} \cdot \vec{\epsilon} \, |\psi_{31m'} \rangle|^2
  \nonumber \\
   & &  \hspace{0.6cm} \Rightarrow \hspace{0.4cm} B_S(\omega) \, \simeq \,
   \frac{1}{n^2} \,  B_{3 \rightarrow n}(\omega)         \label{bb}
\end{eqnarray}
It is worth noting that, because the normalized Rydberg state wavefunctions
scale as $n^{-3/2}$ \cite{gallag}, the Einstein coefficient scales as
$n^{-3}$ and
\begin{equation}
     B_S(\omega) \, \propto \, \frac{1}{n^5} \,\, .
\end{equation}
This result, together with the level density formula of Eq.~(\ref{density}),
brings about the important conclusion that \emph{the absorption probability
  $W_{3n}$ is practically independent of $n$ and of the transverse Ps
  velocity}.

Incidentally it is worth noting that Eq.~(\ref{density}) could be straightforwardly written down with the crude argument that a Rydberg level contains $n^2$ sublevels and the interdistance between $n$ levels is $\Delta E_n$. Anyway, this consideration does not account for the strong sublevel opening and mixing. Within that framework, the above statement on the independence on $n$ of the absorption probability $W_{3n}$ requires a sufficiently wide laser bandwidth (covering several unperturbed  $n$ levels), whereas our calculations indicate that this condition is not required.

The strong energy sub-level mixing leads to a physical system
with a wide energy bandwidth with uniform energy level distribution. As a
consequence, the Doppler effect simply leads to a shift of
the resonant frequency for a particular transition within the bandwidth.
The above ``conservation rule'' supports the consideration that for high--$n$
the Doppler effect gives negligible contribution to the global excitation
dynamics.

Using Eqs.~(\ref{density}) and (\ref{bb}), and following the procedure and the definitions outlined in the Appendix, the absorption probability rate turns out to be:
\begin{eqnarray}
  W_{3n}(t) \, & \simeq & \, \int_{\Delta E_L} d\omega \,
    \frac{I(\omega,t)}{c} \, \frac{B_{3 \rightarrow n}(\omega)}{n^2} \, \rho(\omega) \nonumber \\
   & = &\, \frac{I_L(t)}{c} \, n^3 \, B_{3 \rightarrow n} \,\,
    \frac{\hbar}{13.6 \, \mbox{eV}} \,\, .    \label{excp3}
\end{eqnarray}
Finally, by considering for definiteness linear laser polarization parallel to the magnetic field direction (hence operating with the selection rule $\Delta m = 0$), we obtain the saturation fluence for the second transition:
\begin{equation}\label{f2}
    F_{sat}(3 \rightarrow n) \, \simeq \, \frac{c \times 13.6 \, \mbox{eV}}
    {B_{3 \rightarrow n} \, \hbar \, n^3} \,
    \simeq \, 0.98 \, \mbox{mJ/cm}^2 \,\, ,
\end{equation}
which in fact results approximately a constant in the useful range between $n=20$ and $n=30$. We can evaluate the total energy of the laser pulse needed for saturation of this Rydberg excitation with the same method used in the previous subsection, and the result is $E_{II} = 174~\mu$J.

\subsection{The final two--step excitation and its numerical simulation}

In the above sections we have found the minimum laser energy requirements to
obtain saturation on the two
transitions. However, the real Rydberg excitation is performed with near
simultaneous laser pulses. This is so because of the narrow useful time
window, to cope with a rather large expanding Ps cloud, and the need of
avoiding losses on $n = 3$ excited population due to its non
negligible spontaneous emission. In this conditions the excitation
dynamics involves all the three levels of the two--step transition. If
the laser pulse energies are greater than the saturation fluences, an
overall level population of 33\% is expected for this incoherent excitation \cite{shore}, in the limit case of no losses. This can be confirmed with a dynamical model as follows.

In the previously discussed picture of the problem there is a lack of
information on the exact quantum numbers for the final states of the
transition. Therefore we decided to use a simplified excitation dynamics
model to
obtain an estimate of the high-$n$ state population. We made
dynamical simulations considering transitions from $(n,l,m) = (1,0,0)$ to
the state $(3,1,0)$ and from this state to the final states $(n',2,0)$ and
$(n',0,0)$, assuming linear laser polarization as discussed before. In simulations we have considered the total cross section of the transition from the lower to the upper level band substantially equal to the transition cross section
between the two levels connected by electric dipole selection
rules. This choice is quite usual in problems of this kind
\cite{shore}, and can be inferred from the discussion in the previous
subsection.

The resonant Ps excitation is described with a model of multilevel Bloch
equation system, derived from a complete density matrix formulation
\cite{meystre}, and including for completeness population losses due to
spontaneous decay and photoionization for both excited states. Inserting
photoionization is necessary for a correct description of the dynamics of
Rydberg level population, because this process is in competition with the
Rydberg excitation. A rough estimate of the ionization cross sections can
be made with the method of Ref.~\cite{noordam}. The ionization probability,
proportional to the total energy of the laser pulses, is higher for the sequence $1
\rightarrow 3 \rightarrow $ high--$n$ with respect to the alternative $1
\rightarrow 2 \rightarrow $ high--$n$, but remains limited to the very small
value of 0.3\% .  In particular, in the case of the second excitation the
very long spontaneous emission lifetime and the relatively high ionization
cross section make ionization processes responsible for the overwhelming
majority of the population loss rate.

Since the laser pulses are
substantially incoherent, the light phase in our model is taken as a
``random walk'' with the step equal to the coherence time.
\begin{figure}[!ht] \center
  \includegraphics[width=8.0cm,clip=true]{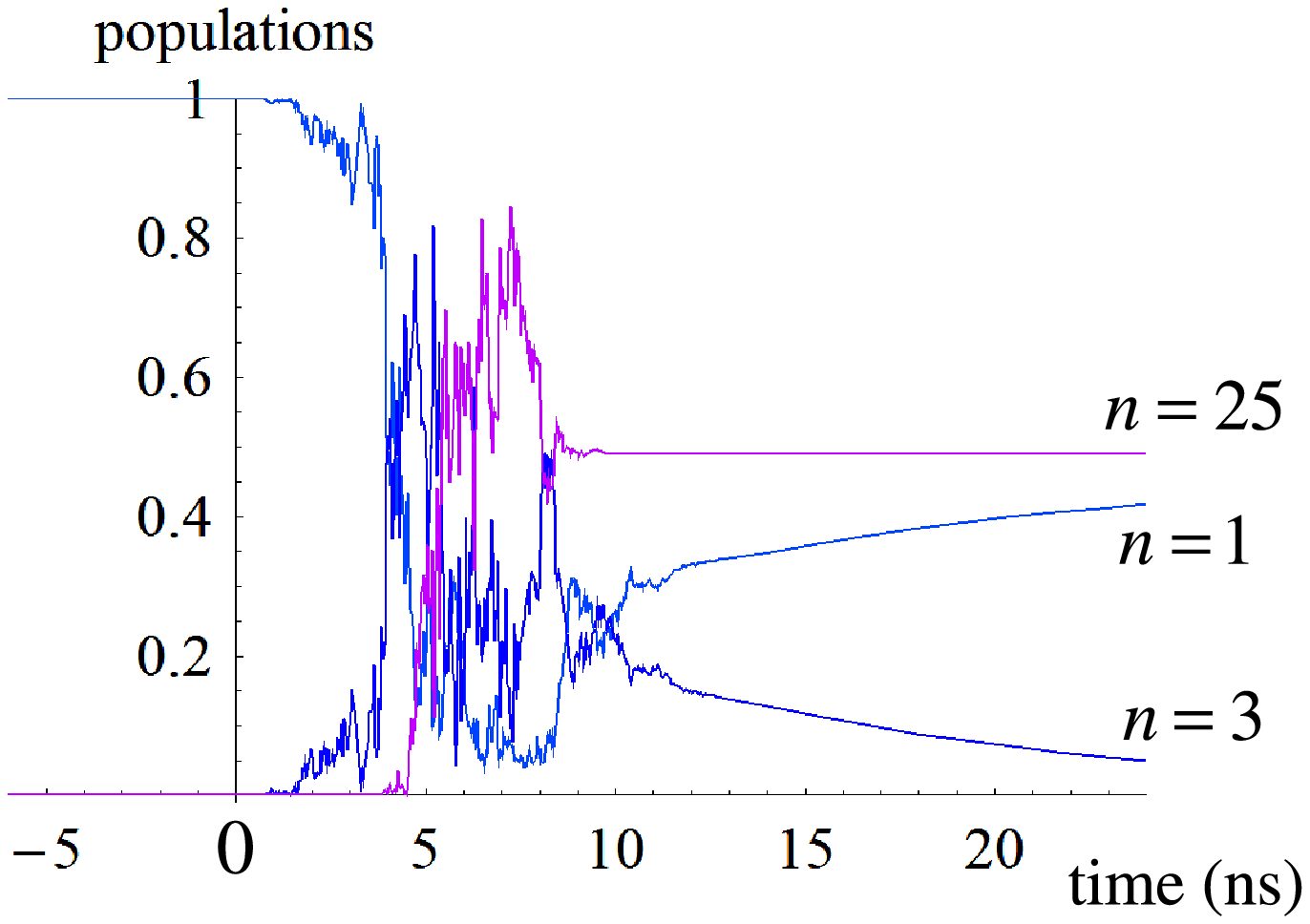}
  \caption{(Color online) Plot of level populations versus time in a single realization of the excitation process $1 \rightarrow 3 \rightarrow 25$.  The characteristics of the two laser pulses are: (1) time length 4 ns, fluence 200 $\mu$J/cm$^2$ and spectral width $\Delta \lambda = 0.045$ nm, (2) time length 2 ns, fluence 2.0 mJ/cm$^2$ and spectral width $\Delta \lambda = 0.72$ nm (two times the Doppler bandwidth), respectively.}
\label{sim}
\end{figure}
Fig. \ref{sim} shows an example of the fractional level populations of Ps as a
function of time when irradiated with two simultaneous laser pulses, the
first one resonant with the transition $1 \rightarrow 3$ and the second one resonant
with the unperturbed transition $3 \rightarrow 25$ (specifically $(1,0,0)
\rightarrow (3,1,0) \rightarrow (25,2,0)$). Both pulses have a fluence $F(t)$
(spectral integrated intensity) slightly greater than twice the saturation
fluence of the relative transitions, to compensate for population loss.
Other characteristics of the pulses are listed in the figure caption.

\begin{figure}[!ht] \center
  \includegraphics[width=7.8cm,clip=true]{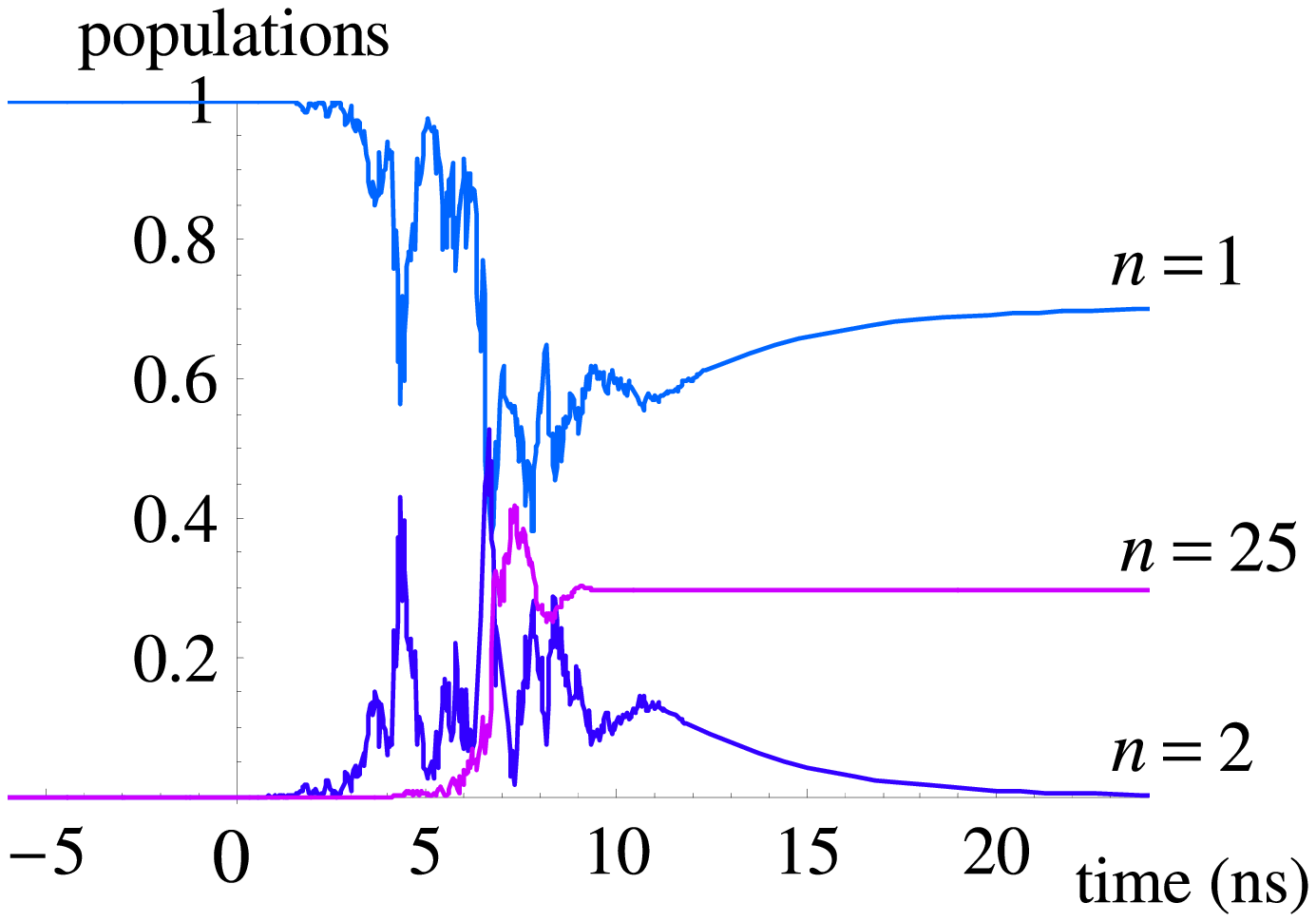}
  \caption{(Color online) Plot of level populations versus time in a single realization of the alternative excitation process $1 \rightarrow 2 \rightarrow 25$.  The characteristics of the two laser pulses are: (1) time length 4 ns, fluence 25.7 $\mu$J/cm$^2$ and spectral width $\Delta \lambda = 0.054$ nm, (2) time length 2 ns, fluence 8.0 mJ/cm$^2$ and spectral width $\Delta \lambda = 0.36$ nm, respectively.}
\label{sim2}
\end{figure}

The final excitation probability for the entire Ps cloud comes from an
averaging process over many simulation outputs.
The calculation shows that a fraction of about 30\% of Ps atoms are excited
to the Rydberg state, and this result does not change by irradiating with
larger laser fluences, or by considering the slightly less effective transition
to the state $(25,0,0)$. As a comparison, a parallel numerical simulation
can be done in the case of the alternative sequence of excitations $1
\rightarrow 2 \rightarrow 25$, using the same rules to determine the required
laser fluences and bandwidth. An example of the fractional level populations of Ps as a function of time is shown in Fig. \ref{sim2}. The pulses maintain the same time length as before, and the other  characteristics are listed in the figure caption. Note that in this case higher energy is required for
the second laser pulse. The fraction of excited Ps atoms results around
24\%, mainly because the intermediate level suffers an increase in
population losses and Doppler bandwidth, which affects the incoherent
excitation dynamics with a reduction in average excitation efficiency.

\section{Conclusions}

Excitation process to high-$n$ levels of Ps having 100 K temperature and set
in a strong magnetic field presents the peculiarity of a complete
re-organization of the energy level structure.  In fact, the motional Stark effect
splits and totally mixes the otherwise degenerate sublevels of Rydberg states
leading to systems with wide energy bandwidth with a uniform energy level
distribution. Therefore, Stark effect totally overcomes the Doppler effect in
determining the bandwidth of the transition. It is important to note that
both effects, Doppler and Stark, depend on the square root of the
temperature, hence one expects transition characteristics not to
change if the Ps cloud can be extracted from the silica converter with lower
average kinetic energy.

The main consequence of the motional Stark effect on the resonant
excitation with laser pulses is that it involves a large number of Rydberg
states of different $n$, whose sublevels are interleaved. This theoretical
result certainly needs an experimental evidence.  Another important
consequence of the Stark effect is that the range of high-$n$ levels on which
one can obtain high excitation efficiency is limited (for example at $n \leq
27$ for $T = 100$ K) because Ps atoms populating part of the $n$-sublevels
are easily ionized by the induced electric field.

Calculations performed in this paper foresee a 30 \% transition efficiency with
a tailored laser system (with $n$ below the limit for Stark ionization as
discussed before). In particular, the degree of freedom of the interaction
bandwidth with the Rydberg level manifold is important to obtain
excited Ps atoms in a definite range of energies, which may be
required for maximum efficiency of the subsequent charge exchange process
with antiprotons. It must be further researched how Ps atoms populating
different n--levels, even if with  nearby energy values, affect the
efficiency of the charge exchange reaction
$Ps^* + \overline{p} \rightarrow \overline{H} + e^-$ .

Simple considerations based on the general theory of radiative transition in
atomic physics are used to derive rules which give the  laser pulse
fluence required for transition saturation. In the first step $1
\rightarrow 3$ a relatively simple modeling is sufficient, being the
bandwidth essentially Doppler dependent. For the second step $3 \rightarrow
n$ more attention must be devoted because of the strong motional Stark mixing
affecting the Rydberg level structure, resulting in the formula (\ref{f2}).

Finally, we stress that the discussed Ps excitation scheme makes a complete
incoherent population transfer, because of the very short coherence length of
the two lasers required for the excitation.  This is reflected in the maximum
excitation efficiency, that cannot be more than 33\% \cite{shore}.
A possible extension of this work towards higher excitation efficiency is the use of
coherent population transfer techniques such as
STIRAP (STimulated Raman Adiabatic Passage) \cite{shore2}, which appears nevertheless challenging because of the complex structure of the Rydberg levels.

\appendix*
\section{Definition of saturation fluence}

The dynamics of atomic incoherent excitation by a laser pulse can be
described by a rate equation model \cite{shore}. Considering for
definiteness the dipole allowed transition $(n,l,m) = (1,0,0)$ to
$(n,l,m) = (3,1,m)$, the rate equation for the high level population
$P_3$ is

\begin{equation}\label{p3}
    \frac{d P_3}{d t} \, = \, - P_3 \, W_{SE} \, - P_3 \, W_{31}(t) \, +
    \, P_1 \, W_{13}(t)
\end{equation}
where $W_{SE}$ is the total spontaneous emission rate, $ W_{13}(t)$
the absorption probability rate given by Eq.~(\ref{excp1}), and
$W_{31}(t)$ the stimulated emission probability rate. Observing that the 10.5 ns lifetime
of the $n = 3$ state is longer than the laser pulse time length which governs the excitation dynamics time scale, for simplicity we discard the spontaneous emission rate, even if the laser pulse duration is not totally negligible compared to it. Therefore it holds $P_1 + P_3=1$ for the lower and upper level populations. Assuming that the transition is ruled by a polarized laser pulse, \emph{i.e.} fixing $\Delta m$, we
have equal probability for photon absorption and stimulated emission,
therefore

\begin{equation}\label{p3bis}
    \frac{d P_3}{d t} \, \simeq \, \left( 1 - 2 P_3 \right) \, W_{13}(t) \,\, .
\end{equation}

The excitation is performed with a laser pulse having a total intensity $I_L(t) = \int d\omega \, I(\omega,t)$, where $I(\omega,t)$ is a time--dependent Gaussian spectral intensity resonant with the transition frequency $\omega_{13}$. By
selecting the laser broadening equal to Doppler broadening (as required in Section 3.1) and using the fact that the absorption coefficients $B_{1 \rightarrow 3}(\omega)$ of Eq.~(\ref{cs1}) is in practice a constant, the rate equation (\ref{p3bis}) can easily be solved obtaining

\begin{equation}\label{psol}
    P_3 (t) \, = \, \frac{1}{2} \, \left[ 1 - e^{-2 F(t)/F_{sat}} \right]
\end{equation}
where

\begin{equation}\label{ft}
    F(t) \, = \, \int_{-\infty}^{\,t} \, dt' \, I_L(t')
\end{equation}
is the laser pulse fluence, and
\begin{equation}\label{fsat}
    F_{sat} (1 \rightarrow 3)\, = \, \frac{c \, \sqrt{2}}{B_{1 \rightarrow 3} \, g_D(0)}
\end{equation}
is the saturation fluence. This parameter characterizes the population
dynamics: from Eq.~(\ref{psol}) it is clear that when $F(t) = F_{sat}$
we have 43\% of the atoms in the excited state. The maximum
excitation, \emph{i.e.} the saturation level of the transition, reaches
50\% with high enough pulse energy .

In the case of the second transition, from $(3,1,m)$ to Rydberg levels, a similar calculation is performed without reference to the Doppler bandwidth and considering that
$B_{3 \rightarrow n}(\omega)$ is essentially a constant over a wide frequency range. The saturation fluence is then given by Eq.~(\ref{f2}).

\begin{acknowledgments}
We acknowledge a useful discussion with S.Hogan of the ETH--Zurich and with C.Drag of the Laboratoire Aim\'e Cotton.
\end{acknowledgments}


\begin{thebibliography}{50}
\bibitem{note1} This result can be easily derived from eq.~(\ref{estark}) used in the extended discussion over the second transition.
\end{thebibliography}
\end{document}